\documentclass[prd,twocolumn,superscriptaddress,altaffilletter,showpacs]{revtex4}
\usepackage{amssymb,amsmath}

\begin{document}
\title{
Thick Brane Isotropization in a Generalized 5D Anisotropic Standing Wave Braneworld Model
}

\author{Merab Gogberashvili} \email{gogber@gmail.com}
\affiliation{Andronikashvili Institute of Physics, 6 Tamarashvili St., Tbilisi 0177, Georgia and\\
Javakhishvili State University, 3 Chavchavadze Ave., Tbilisi 0128, Georgia}

\author{Alfredo Herrera--Aguilar} \email{aha@fis.unam.mx}
\affiliation{Instituto de Ciencias F\'{\i}sicas, Universidad Nacional Aut\'onoma de M\'exico,\\
Apdo. Postal 48-3, CP 62251, Cuernavaca, Morelos, M\'{e}xico}
\affiliation{Instituto de F\'{\i}sica y Matem\'{a}ticas, Universidad Michoacana de San Nicol\'as de \\
Hidalgo, Edificio C--3, Ciudad Universitaria, CP 58040, Morelia, Michoac\'{a}n, M\'{e}xico}

\author{Dagoberto Malag\'on--Morej\'on} \email{malagon@fis.unam.mx}
\affiliation{Instituto de Ciencias F\'{\i}sicas, Universidad Nacional Aut\'onoma de M\'exico,\\
Apdo. Postal 48-3, CP 62251, Cuernavaca, Morelos, M\'{e}xico}

\author{Refugio Rigel Mora--Luna} \email{rigel@ifm.umich.mx}
\affiliation{Instituto de F\'{\i}sica y Matem\'{a}ticas, Universidad Michoacana de San Nicol\'as de \\
Hidalgo, Edificio C--3, Ciudad Universitaria, CP 58040, Morelia, Michoac\'{a}n, M\'{e}xico}

\author{Ulises Nucamendi} \email{ulises@ifm.umich.mx}
\affiliation{Instituto de F\'{\i}sica y Matem\'{a}ticas, Universidad Michoacana de San Nicol\'as de \\
Hidalgo, Edificio C--3, Ciudad Universitaria, CP 58040, Morelia, Michoac\'{a}n, M\'{e}xico}

\date{\today}

\begin{abstract}
We study a smooth cosmological solution within a generalized 5D standing wave braneworld modeled by gravity and 
a phantom scalar field. In this model the 3-brane is anisotropically warped along its spatial dimensions and 
contains a novel time-dependent scale factor that multiplies the anisotropic spatial interval of the 5D metric, 
a fact that allows us to study cosmological effects. By explicitly solving the bulk field equations we found a 
natural mechanism which isotropizes the braneworld for a wide class of natural initial conditions. We are able 
to give a physical interpretation of the anisotropic dissipation: as the anisotropic energy of the 3-brane 
rapidly leaks into the bulk through the nontrivial components of the projected to the brane non-local Weyl tensor, 
the bulk becomes less isotropic. At the same time, under the action of the 4D cosmological constant, the anisotropic 
braneworld super-exponentially isotropizes by itself, rendering a 3-brane with de Sitter symmetry embedded in a 5D 
de Sitter space-time, while the phantom scalar field exponentially vanishes.
\end{abstract}


\maketitle


\section{Introduction}

Braneworld models involving large extra dimensions have been very useful in addressing several open questions in high energy physics \cite{Hi,brane}, and astrophysics and cosmology (for reviews see \cite{reviews,maartenskoyama,thickreview}). For more than a decade it is clear that braneworlds possess a set of appealing features that encourage one to develop this research line further. Namely,
braneworld models recast the hierarchy problem into a higher dimensional viewpoint, reveal a geometrical mechanism of dimensional reduction supported by a curved extra dimension, can trap various matter fields on the brane (including gauge bosons in some cases), take into account the self-gravity on the brane through its tension, give a realization of the AdS/CFT correspondence to lowest order, allow one to address the smallness of the cosmological constant, provide several cosmological backgrounds with consistent dynamics that lead to General Relativity results under suitable restrictions on their parameters, etc. Most of the braneworlds were realized as time independent field configurations. However, mostly within the framework of cosmological studies, there have been proposed different braneworld models that employ time-dependent metrics and matter fields (if any), as well as branes with tensions varying in time \cite{Cosmbranes,S,kasnerads,anisotbwinfl,bwcosmolanisot,bcosmoanisotbulk,bwisotropization,GK,GHMMcosmology,vartensions}.

It is widely believed that our Universe started its evolution from a highly anisotropic state which lead to a highly symmetric state that we observe nowadays (according to the recent WMAP data CMB is isotropic with an accuracy of $10^{-5}$ \cite{wmap}). An interesting cosmological question concerns the dynamical mechanism that led the Universe to shed almost all of its anisotropy along time till the present epoch which is highly isotropic. A very popular mechanism among cosmologists is inflation: the Universe diluted away almost all of its anisotropy during a period of accelerating expansion in an early epoch. Of course, there are several other mechanisms that tried to explain this important phenomenon (see \cite{bwisotropization} and references therein).

When studying braneworlds generated by anisotropic backgrounds, one must have a complete solution to the bulk and brane field equations in order to consistently analyze the cosmological dynamics on the brane. Such a solution involves the knowledge of the bulk Weyl tensor which is felt on the brane by means of its projection \cite{SMS}. It turns out that this is not an easy task to solve \cite{bwcosmolanisot} and up to now, there are a very few cosmological anisotropic braneworlds that present a complete solution \cite{kasnerads,anisotbwinfl,bwcosmolanisot,bcosmoanisotbulk,bwisotropization,GK,GHMMcosmology}. In this context, there are many studies reported in the literature that made several assumptions about the Weyl term in the absence of knowledge of the full solution to the bulk metric \cite{assump1,assump2}. The main difficulty here consists in finding anisotropic generalizations of the $AdS_5$ space-time that are necessarily non-conformally flat and incorporate anisotropy on the cosmological brane. In this sense, it is relevant to propose new braneworld generalizations that attempt to describe more realistic cosmological braneworld models, or explore other aspects of higher-dimensional gravity which were not addressed till now by the known brane models.

Recently a braneworld generated by 5D anisotropic standing gravitational waves minimally coupled to a phantom-like scalar field was proposed \cite{Gog1,ghamm}. Usually in 4D cosmology phantom fields (mostly scalars \cite{Cal}) are used to model dark energy \cite{Dark}. It seems that WMAP data combined with either Supernova or Baryonic Acoustic Oscillations prefer the phantom model of dark energy \cite{wmap}. The motivation to introduce a phantom scalar fields in the 5D model \cite{Gog1,ghamm} was to provide an alternative mechanism for localizing 4D gravity as well as for trapping matter fields \cite{gmmscalar}, including gauge bosons \cite{gmmboson}, which usually are not localized on thin braneworlds. In this model the phantom scalar field is present only in the bulk and does not couple to ordinary matter (to avoid the well-known problems of stability which occur with ghost fields). Notwithstanding, the metric and scalar fluctuations are coupled in this system and the problem of stability could persist in principle at the quantum level. 
In Section III we discuss on this issue in more detail.

In this paper we consider a thick version of the braneworld model \cite{Gog1,ghamm} and modify the metric {\it ansatz} 
by adding a time-dependent scale factor that multiplies the spatial sector of the anisotropic 3-brane in the 5D metric 
in order to be able to study cosmology and, in particular, a possible isotropization mechanism for the early universe. 
Thus, in comparison to \cite{Gog1,ghamm}, in this model we smooth out the warp factor and incorporate a scale factor 
in the metric. The first generalization allows us to dispense with the use of Dirac delta functions and their 
respective singularities in the model since the time-dependent scalar field acts as a source in complete agreement 
with the fact that anisotropic metrics on the brane are not consistent with static bulks 
\cite{bcosmoanisotbulk,bwisotropization}, whereas the second generalization permits us to study new cosmological 
effects that are not present in the original version of the model.

Further, by exactly solving the 5D Einstein field equations, the bulk cosmological constant turns out to be positive, 
realizing a de Sitter braneworld instead of an Anti de Sitter one. Moreover, by leaving the found solution evolve in 
time, the metric isotropizes and the corresponding scalar field exponentially disappears, leading to a recently 
proposed de Sitter braneworld model purely generated by curvature, i.e., by an interplay between the 4D and 5D 
cosmological constants \cite{cuco}. It turns out that within this regular braneworld one is able to model the thick 
brane in a completely geometrical way, avoiding at all the use of scalar matter. In this model the effective 4D 
Planck mass is finite (and also depends on the Hubble parameter of the metric), the 4D gravity can be localized and 
the corrections to Newton's law have already been computed. As an extra bonus, a mass gap is displayed in the gravity 
spectrum of Kaluza-Klein (KK) excitations (a fact that fixes the energy scale at which these massive fluctuations can 
be excited and enables us to avoid difficulties when analyzing the traces of ultra-light KK excitations) without 
developing naked singularities as in the case of scalar thick brane configurations \cite{hammmln}. Another 
interesting feature of this isotropic thick braneworld model is that its 4D cosmological constant can be made as 
small as one desires without the need of fine tuning the bulk cosmological constant with the brane tension, as it 
happens in the Randall-Sundrum type models \cite{ghlmm}.


\section{The model and the complete solution}

We start with a 5D action which describes gravity coupled to a non-self-interacting scalar phantom-like field \cite{Cal,phantom}, which depends on time and propagates in the bulk:
\begin{equation} \label{action}
S_b = \int d^5x \sqrt{g}\left[\frac{1}{16 \pi G} \left( R - 2\Lambda \right) +
\frac 12\left(\nabla \phi\right)^2\right]~,
\end{equation}
where $G$ and $\Lambda$ are 5D Newton and cosmological constants, respectively. The Einstein equations for the action (\ref{action}) are written in the form:
\begin{equation} \label{field-eqns1}
R_{\mu\nu} = T_{\mu\nu} - \frac 13 g_{\mu\nu} T + \frac 23 g_{\mu\nu} \Lambda =
- \partial_\mu\sigma \partial_\nu\sigma + \frac 23 g_{\mu\nu} \Lambda~,
\end{equation}
where Greek indices refer to 5D and we have redefined the scalar field as $\sigma = \sqrt{8\pi G}\phi$ in order to absorb the 5D gravitational constant.

In order to study the braneworld isotropization mechanism we intend to present in this article, we can start from a 
generalization of the metric {\it ansatz} of \cite{Gog1,ghamm} consistent with the fact that the fields of the setup 
possess both temporal and extra coordinate dependence. Thus, we shall begin by considering the following metric:
\begin{eqnarray}
ds^2 = e^{2A(t,w)}\left(-dt^2+\tilde{g}_{mn}(t,w)dx^mdx^n+dw^2\right)~,
\label{metricTu}
\end{eqnarray}
where the warp factor $A=A(t,w)$ and the 3D metric $\tilde{g}_{mn}(t,w)$ ($m,n=1,2,3$) are arbitrary functions.
Note that without loss of generality we can express $\tilde{g}_{mn}(t,w)=a^2(t){g}_{mn}(t,w)$ and thus,
we can further write the metric (\ref{metricTu}) as:
\begin{eqnarray}
ds^2 &=& e^{2A(t,w)}\left(-dt^2+a^2(t){g}_{mn}(t,w)dx^mdx^n\right)\nonumber\\
&+& e^{2A(t,w)}dw^2~.
\label{metrictu}
\end{eqnarray}
Motivated by \cite{bdel} we further require the condition $\dot A=0$ which corresponds to the case when the radion 
is not dynamical in our model. This condition was used in \cite{bdel} for solving the Einstein's equations with a 
general metric {\it ansatz} under the assumption of a maximally symmetric 3D space, allowing the authors of 
these articles to get an interesting general solution with no separation of variables. This fact implies that 
the warp factor $A(w)$ is a function of the extra coordinate alone. 

We shall further consider that 
\begin{eqnarray}
{g}_{mn}(t,w)=\mbox{\rm diag}\left(e^{u(t,w)},e^{u(t,w)},e^{-2u(t,w)}\right)
\label{metric3D}
\end{eqnarray}
in order to use the 3D spatial metric of \cite{Gog1,ghamm}. Here we have set $g_{11}=g_{22}$ for simplicity since 
under this equality the spatial volume of our metric is still anisotropic.
The {\it ansatz} (\ref{metric3D}) seems very artificial but it is necessary in order for us to be able to obtain 
exact solutions, since when $\mbox{\rm det}g_{mn}=1$ (a constant in general), the Einstein and Klein-Gordon equations 
yield decoupled differential equations that can further be analytically solved. This assumption was used in order to 
construct a {\it standing wave} along the fifth dimension in the braneworld model proposed in \cite{Gog1,ghamm}.

Thus, we arrive at the following 5D interval
\begin{eqnarray}
ds^2 &=& e^{2A(w)}\left\{-dt^2+a^2(t)\left[e^u\left(dx^2+dy^2\right)+e^{-2u}dz^2\right]\right. \nonumber\\
&+& \left. dw^2\right\}~.
\label{metricA}
\end{eqnarray}
This is the metric {\it ansatz} we shall work with in this paper in order to study the above mentioned isotropization 
mechanism. We see that $a(t)$ plays the role of a cosmological scale factor, 
$u(t,w)$ describes time-dependent gravitational warping and the warp factor $A(w)$ is an arbitrary function of the 
extra coordinate $w$, allowing for smooth solutions, and models a thick brane configuration in the spirit of 
\cite{gremm}. 

This metric generalizes straightforwardly the thin brane metric {\it ansatz} of \cite{Gog1,ghamm} 
to the case when the warp function $A(w)$ does not involve the module of the fifth dimension, i.e. $A(w) \neq A(|w|)$ 
(a fact that smoothes out the field configurations avoiding the singularities generated by the Dirac delta sources of 
the setup), and a scale factor $a(t)$ multiplies the 3D anisotropic spatial sector of the metric, introducing 
cosmology apart from the bulk gravitational waves of the original model, where the 3-brane possesses anisotropic 
oscillations which send a wave into the bulk as in \cite{gms}.

Thus, in our model the brane is anisotropically warped along the spatial coordinates $x, y, z$ through the factor $e^{u(t,w)}$, depending on time $t$ and the extra coordinate $w$. Moreover, these warped spatial coordinates are multiplied by a scale factor $a(t)$ that allows them to evolve in time (expanding or contracting) as an extra feature. This model also joins previously constructed anisotropic braneworlds \cite{kasnerads,anisotbwinfl,bwcosmolanisot,bcosmoanisotbulk,bwisotropization,GK,GHMMcosmology} which approached several cosmological issues, like anisotropy dissipation during inflation \cite{anisotbwinfl}, braneworld isotropization with the aid of magnetic fields \cite{bwisotropization} and localization of test particles \cite{GK,gmmscalar,gmmboson}. Moreover, as a general feature it has been established that anisotropic metrics on the brane cannot be supported by static bulks \cite{bcosmoanisotbulk,bwisotropization}.

On the other hand, the phantom-like scalar field $\sigma(t,w)$ obeys the Klein-Gordon equation on the background (\ref{metricA}):
\begin{eqnarray} \label{sigma}
\frac{1}{\sqrt{-g}}~\partial_\mu \left(\sqrt{-g} g^{\mu\nu}\partial_\nu \sigma\right) = \nonumber \\
= \ddot \sigma + \frac {3\dot a}{a}~ \dot\sigma - \sigma'' - 3 A' \sigma' = 0 ~,
\end{eqnarray}
where overdots and primes stand for derivatives with respect to time and the extra coordinate, respectively.

Under the {\it ansatz} (\ref{metricA}) we can write the Einstein equations in the form:
\begin{eqnarray} \label{field-eqns}
R_{tt} &=& -\frac 32 \dot u^2 - 3\frac{\ddot a}{a}+ A'' + 3A'^2 =
- \frac 23 e^{2A}\Lambda - \dot\sigma^2~, \nonumber \\
R_{xx} &=& R_{yy} = a^2e^{u} \left[\frac 12 \left(\ddot u + 3\frac{\dot a}{a}\dot u - u'' - 3A'u'\right)
+ \right. \nonumber\\
&+& \left. \frac{\ddot a}{a} + 2\left(\frac{\dot a}{a}\right)^2 - A''- 3A'^2\right] =
\frac 23 e^{2A+u} a^2 \Lambda  ~, \nonumber \\
R_{zz} &=& a^2e^{-2u}\left[ -\ddot u-3\frac{\dot a}{a}\dot u + u''+ 3A'u' + \frac{\ddot a}{a}+ \right. \\
&+& \left. 2\left(\frac{\dot a}{a}\right)^2 - A''- 3A'^2\right] = \frac 23 a^2 e^{2A-2u} \Lambda~, \nonumber \\
R_{ww} &=& - \frac 32 u'^2 - 4A'' = \frac 23 e^{2A}\Lambda - \sigma'^2~, \nonumber\\
R_{tw} &=& -\frac 32\dot u u' = - \dot\sigma \sigma' ~. \nonumber
\end{eqnarray}

From the $tw$-component of the Einstein equations (\ref{field-eqns}), it follows that the fields $\sigma$ and $u$ are related by
\begin{equation} \label{sigma=u}
\sigma (t,r) = \sqrt{\frac 32}~u(t,r)~,
\end{equation}
up to a meaningless additive constant. The same relationship was obtained in \cite{Gog1,ghamm} and it holds even in 
the presence of an arbitrary self-interaction potential of the scalar field, evidencing the need of a phantom scalar 
in the setup, instead of a canonical scalar field. Another way of looking at this fact is that the quadratic relation 
(\ref{sigma=u}) between $u$ and $\sigma$ fixes the phantom nature of the scalar field regardless of a self-interaction 
potential. If one would require a standard scalar field by changing the sign of the kinetic term in the action, then 
the metric would necessarily become complex.

We further take into account the Klein-Gordon equation (\ref{sigma}) in the Einstein equations (\ref{field-eqns}), and combine the $tt$- and $xx$-components to obtain a single relevant equation for the scale factor:
\begin{equation} \label{eqna}
\frac{\ddot a}{a} - \left(\frac{\dot a}{a}\right)^2 = 0~,
\end{equation}
which has a solution of the form:
\begin{equation} \label{a}
a(t) = a_0 e^{Ht}~,
\end{equation}
where $a_0$ is a constant that can be absorbed in a redefinition of the spatial coordinates.

Once we know the solution for the scale factor, we further substitute its expression back into the Einstein equations, and combine the $xx$- and $ww$-components in order to get an equation for the warp factor $A(w)$:
\begin{equation} \label{eqnA}
A''- A'^2 + H^2 = 0~.
\end{equation}
This equation has the solution:
\begin{equation} \label{A}
A(w)= \ln \left[\frac Hb \mbox{\rm sech}\left[ H(w-w_0) \right]\right]~,
\end{equation}
where the integration constant $b$ is related to the 5D cosmological constant as follows:
\begin{equation} \label{b=Lambda}
\Lambda = 6b^2~,
\end{equation}
which reveals the de Sitter nature of the 5D bulk space-time.

On the other hand, after taking into account the solution for the scale factor (\ref{a}), and assuming the following {\it ansatz} for the scalar field:
\begin{equation} \label{separation}
\sigma(t,w) \sim u(t,w) = \epsilon(t) \chi(w) e^{-\frac 32 A}~,
\end{equation}
the Klein-Gordon equation reduces to a couple of ordinary differential equations:
\begin{eqnarray}\label{chi}
\chi'' - \left(\frac 32 A''+ \frac 94 A'^2 - \Omega^2 \right) \chi = 0~, \\
\label{e} \ddot \epsilon + 3H\dot \epsilon + \Omega^2 \epsilon = 0~,
\end{eqnarray}
where $\Omega$ is an arbitrary constant.

The equation (\ref{chi}) can be regarded as a Schr\"odinger equation with the analog quantum mechanics potential:
\begin{equation}
V_{QM} = \frac 32 A''+\frac 94 A'^2~.
\end{equation}
Using the solution for the warp factor (\ref{A}) we see that it represents a modified P\"oschl-Teller potential of the form:
\begin{eqnarray}\label{VQM}
V_{QM}(w)=\frac 94 H^2 - \frac {15}{4} H^2 {\rm sech}^2\left[ H(w-w_0) \right]~.
\end{eqnarray}
Thus, the differential equation for $\chi$ (\ref{chi}) turns out to be a known eigenvalue problem with a mixed spectrum (see \cite{massgap}, for instance). Namely, there are two bound states: a ground state with $\Omega=0$ and another one with $\Omega=\sqrt{2}H$, separated by a gap determined by the asymptotic value:
\begin{equation}
V_{QM} (\infty)=\frac 94 H^{2}~,
\end{equation}
as well as a continuum of KK states starting at $\Omega= 3H/2$.

In fact, the equation for $\chi$ is precisely the eigenvalue equation for the transverse traceless KK metric fluctuations and massless scalar perturbations of the recently proposed de Sitter braneworld model \cite{cuco,ghlmm}, where there is a mass gap in the graviton and scalar mass spectra of KK perturbations.

Thus, the equation for the relevant function $\chi$ possesses a general solution:
\begin{eqnarray} \label{chisol}
\chi(w) = C_1 P^\mu_{3/2}\left[\tanh\left(H(w-w_0)\right)\right] + \nonumber \\
+ C_2 Q^\mu_{3/2}\left[\tanh\left(H(w-w_0)\right)\right]~,
\end{eqnarray}
where $P^{\mu}_{3/2}$ and $Q^{\mu}_{3/2}$ are associated Legendre functions of first and second kind, respectively, degree $\nu=3/2$ and order
\begin{equation}
\mu = \sqrt{\frac 94-\frac{\Omega^2}{H^2}}~.
\end{equation}
Thus, the first discrete state is the ground state with $\Omega=0$, order $\mu=3/2$ and
\begin{equation}
E_0 = -\frac 94 H^{2}~.
\end{equation}
The explicit expression of this zero mode is given by:
\begin{equation}\label{O0}
\chi_0 (w) = c_0 {\rm sech}^{3/2}\left[ H(w-w_0) \right]~,
\end{equation}
where $c_0$ is a normalization constant. On the other hand, the second bound state corresponds to an excited mode with $\Omega=\sqrt{2}H$, order $\mu = 1/2$ and energy
\begin{equation}
E_1 = -\frac 14 H^2~,
\end{equation}
and has the following form:
\begin{equation}
\chi_1 (w) = c_1 \sinh\left[H(w-w_0)\right]{\rm sech}^{3/2}\left[ H(w-w_0) \right]~,
\end{equation}
where $c_1$ also is a normalization constant. Finally, for the continuum of KK massive modes the order becomes purely imaginary $\mu=i\rho=i\sqrt{\Omega^2/H^2 - 9/4}$.

Turning back to the equation for $\epsilon(t)$ (\ref{e}) we see that it is the equation for a damped oscillator which has three different solutions depending on the kind of damping, i.e. on the relation between the constants $H$ and $\Omega$: a). under-damping ($\Omega^2 > 9H^2/4$), b). critical damping ($\Omega^2 = 9H^2/4$), and c). over-damping ($\Omega^2 < 9H^2/4$).

a). The solution for the under-damped case reads:
\begin{equation}\label{undere}
\epsilon(t)=Ce^{-\frac 32 Ht} \sin\left(\omega t + \delta\right)
\end{equation}
where the constant $C$ denotes the oscillations amplitude, the parameter
\begin{equation}
\omega=\sqrt{\Omega^2 - \frac 94 H^2}~,
\end{equation}
is the un-damped frequency of the oscillations and $\delta$ is an arbitrary phase constant. The constants $C$ and 
$\delta$ are determined by initial conditions. We can easily see that these oscillations will exponentially decay to 
zero with time, a fact that translates into a vanishing metric function $u\rightarrow 0$, which leads in turn to an 
isotropic 5D metric with a 3-brane with de Sitter symmetry \cite{cuco}.

b). The solution for the critical damping case reads:
\begin{equation}\label{crite}
\epsilon(t) = e^{-\frac 32 Ht}\left(\alpha t + \beta\right)~,
\end{equation}
where $\alpha$ and $\beta$ are arbitrary constants determined by initial conditions. In this case we observe the same effect as in the under-damped one: the amplitude of the function $\epsilon$ exponentially vanishes with time, making the metric function $u$ disappear and yielding a de Sitter 3-brane embedded in an isotropic 5D de Sitter space-time.

c). Finally, the over-damped case possesses the following solution:
\begin{equation}\label{overe}
\epsilon(t) = e^{-\frac 32 Ht}\left(\alpha e^{\tilde\omega t} + \beta e^{-\tilde\omega t}\right)~,
\end{equation}
where
\begin{equation}
\tilde\omega=\sqrt{\frac 94 H^2-\Omega^2}~, \label{omegat}
\end{equation}
and $\alpha$ and $\beta$ are arbitrary constants determined by initial conditions. Once again we see that the metric function $u$ exponentially decays in time, leading to a completely isotropic 5D metric with a de Sitter 3-brane embedded in it and realizing a very natural isotropization mechanism.

Thereby, we see that in the general case, the solution for the time evolution of the metric function $u(t,w)$ 
expressed by (\ref{separation}) super-exponentially yields an isotropic 5D metric of the form:
\begin{eqnarray} \label{isotropicmetric}
ds^2 &=& e^{2A(w)}\left[-dt^2 + dw^2 + \right. \\
&+& \left. a^2(t)\left(dx^2 + dy^2 + dz^2 \right) + \right]~, \nonumber
\end{eqnarray}
which possesses an embedded 3-brane with de Sitter symmetry (\ref{a}) under a wide class of initial conditions. 
Namely, for an arbitrary $t_i>0$, the anisotropic metric will super-exponentially evolve to an isotropic 5D metric 
since all the solutions for $\epsilon(t)$ exponentially vanish in time. Moreover, this effect takes place for any 
values of the integration constants of the solution to the braneworld model $C$, $\alpha$, $\beta$ and $\delta$. It is 
worth noticing that together with the function $u$, the scalar field also exponentially disappears as a consequence of 
their proportionality (\ref{sigma=u}), rendering a completely geometric de Sitter thick brane with very appealing 
features \cite{cuco}.

We now proceed to briefly review this purely geometric isotropic braneworld model. It arises from the action
\begin{equation} \label{geometricaction}
S = 2M^3 \int d^5x \sqrt{g}\left( R - 2\Lambda \right)~,
\end{equation}
where we have redefined the gravitational coupling constant as $2M^3=\frac{1}{16 \pi G}$, and is generated by the pure positive curvature of both 5D and effective 4D cosmological constants under the {\it ansatz} (\ref{isotropicmetric}). In order to extract an effective 4D Einstein-Hilbert action on the brane from the bulk action (\ref{geometricaction}) we express it as:
\begin{eqnarray}
S_{eff}\supset \int d^4x\sqrt{-\bar g}~2M^3\int^{\infty}_{-\infty}dw \left\{e^{3A}\bar R + \right.\\
+ \left. 4H^2e^{3A}\left[5~{\rm sech}^2(Hw)-3\right] - 2e^{5A}\Lambda_5 \right\}~, \label{Seff}
\end{eqnarray}
where $\bar R$ was computed with the barred metric $\bar g_{MN}=e^{-2A}g_{MN}$ and we have taken into account the form of the warp factor (\ref{A}). We further integrate over the extra coordinate $w$ and compare the result to the 4D Einstein-Hilbert action on the brane,
\begin{equation}
S_{brane} = 2M^2_{pl}\int d^4 x \sqrt{-^4g} \label{4Daction} \left(^4 R - 2\Lambda_4\right)~.
\end{equation}
In order to derive the scale of the effective 4D gravitational interactions we focus on the curvature term and get:
\begin{eqnarray}
M^2_{Pl} = M^3\int^{\infty}_{-\infty} dw e^{3A(w)} = \nonumber \\
= \frac{M^3H^3}{b^3} \int^{\infty}_{-\infty} dw~{\rm sech}^3(Hw) = \frac{\pi M^3H^2}{2b^3}~,
\end{eqnarray}
which is finite, as it should be for a well defined 4D theory, and explicitly depends on both 5D and 4D cosmological constants since the second and third terms in (\ref{Seff}) contribute to the definition of the effective 4D cosmological constant $\Lambda_4$:
\begin{equation}
\Lambda_4 = 3H^2~.
\end{equation}
In contrast with the Randall-Sundrum thin braneworld model, this cosmological constant can be made as small as one desires without fine tuning it to zero with the aid of the bulk cosmological constant (\ref{b=Lambda}) and the tension of the brane since here we do not have junction conditions. Moreover, both cosmological constants are defined through different independent parameters (see \cite{ghlmm} for details).

In order to study the localization of 4D gravity and further compute the corrections to Newton's law in this model we need to first perturb the isotropic metric,
\begin{equation}
ds^2 = e^{2A(w)}\left[dw^{2}+\left( g_{\mu \nu}+h_{\mu \nu}(x,w) \right)dx^{\mu}dx^{\nu}\right]~,
\label{metricz}
\end{equation}
in the axial gauge $h_{5M}=0$, compute the Einstein's equations to first order for the transverse traceless sector $h^{\mu}_{\mu}=\partial^{\mu}h_{\mu\nu}=0$ of the tensorial fluctuations, further denoted by $\bar h_{\mu \nu}$, and solve the relevant bulk equations in order to get explicit expressions for them. After this work is done, one can calculate the corrections to the Newtonian gravitational potential due to the KK massive modes of these fluctuations in the thin brane limit by making use of a simple formula. Thus, if we consider the variable separation $\bar h_{\mu\nu} = e^{-\frac 32 A(w)}\Psi(w)g(x)_{\mu\nu}$, the transverse traceless modes of the metric fluctuations obey the following Schr\"odinger-like equation along the fifth dimension:
\begin{equation}
\left(-\partial^2_w+\frac 94 A'^2 + \frac 32 A'' - m^2 \right)\Psi(w)=0~, \label{Scheqn}
\end{equation}
where one can identify an analog quantum mechanical potential $V_{QM}$:
\begin{equation}
V_{QM}=\frac 94 A'^2+\frac 32 A'' = -\frac{15}{4} H^2 {\rm sech}^2(Hw)+\frac 94 H^2~,
\end{equation}
which is of  modified P\"oschl-Teller type. This potential ensures the existence of a mass gap in the spectrum determined by its asymptotic value $9H^2/4$ and possesses two bound states. One of them represents a massless graviton which is localized in the brane and is responsible for recovering 4D gravity on the brane with the {\it ansatz} (\ref{metricz}) with no tachyonic modes at all \cite{cuco}. The second one is an excited KK massive mode.

In (\ref{Scheqn}) the parameter $m^2$ corresponds to the mass in a 4D de Sitter space-time in accordance with the relevant 4D equation:
\begin{eqnarray}
\left(-\partial^2_t - 3H\partial_t + e^{-2Ht} \nabla^2 - 2H^2 \right)g(x)_{\mu\nu}= \nonumber \\
= -m^2 g(x)_{\mu\nu}~,
\end{eqnarray}
with $\nabla^2$ being the Laplacian in 3D flat space.

The general solution to the Schr\"odinger equation (\ref{Scheqn}) is given by:
\begin{equation}
\Psi(z) = C_1 P^\mu_{\frac 32}\left(\tanh(Hw)\right) + C_2 Q^\mu_{\frac 32}\left(\tanh(Hw)\right)~,
\end{equation}
where $P^\mu_{\frac 32}$ and $Q^\mu_{\frac 32}$ are associated Legendre functions of first and second kind, respectively, of degree $\nu=3/2$ and order $\mu=\sqrt{9/4 - m^2 /H^2}$.

The continuum of KK massive modes starts from $m\geq 3H/2$ and is described by the following eigenfunctions with imaginary order $\mu = i\rho = \sqrt{m^2/H^2 - 9/4}$~,
\begin{equation}
\Psi_m (w) = \sum_\pm C_\pm P^{\pm i\rho}_{\frac 32}\left( \tanh(Hw) \right)~,
\label{massmodes}
\end{equation}
which asymptotically behave as plane waves.

The massive modes contribute to Newton's law with small corrections coming from the fifth dimension. In order to compute them we consider the thin brane limit $H\longmapsto \infty$ in the resulting isotropic model, locate two probe masses $M_1$ and $M_2$ in the center of the brane in the transverse direction $r$ and compute their full interaction potential. Thus, the corrections to the potential generated by the massive gravitons can be expressed as follows
\begin{eqnarray}
U(r) \sim \frac{M_1M_2}{r} \left(G_4 + \int_{m_0}^{\infty}dm \frac {e^{-mr}}{M^3} \left|\Psi^{m}_\mu(w_0)\right|^{2} + \right. \nonumber \\
+\left. \frac {e^{-m_1r}}{M^3} \left|\Psi_1(w_0)\right|^{2} \right) =
\frac{M_1M_2}{r}\left(G_4 + \triangle G_4\right)~,
\end{eqnarray}
where the brane is located at $w = w_{0}$, $G_4$ is now the gravitational coupling constant in 4D, $\Psi_1(w_{0})$ represents the first excited state, it does not contribute in the thin brane limit that we are considering here since it is an odd function; $\Psi_\mu(w_{0})$ denotes the continuous KK massive states that need to be integrated over their masses. After some simple procedure we get the corrections to Newton's law due to the continuum of KK massive modes (see \cite{massgap} and references therein for details):
\begin{equation}
\triangle G_4\sim \frac{1}{\left| \Gamma(-\frac 14)\Gamma(\frac 74)
\right|^2}\frac{e^{-\frac 32 Hr}}{M^3r}\left[1 + O \left(\frac{1}{Hr}\right)\right],
\end{equation}
which exponentially decay for large enough $Hr$, indicating its small character.

Finally, in order to understand in detail what happens to the anisotropy of the initial scalar-tensor system and give a consistent physical interpretation to the isotropization process, it is necessary to compute the Weyl tensor and its projection to the brane using, for example, the formalism of \cite{SMS}. However, it is clear that the inflationary process that isotropizes the metric is due to the presence of the 4D cosmological constant (since it is equal to $\Lambda_4 = 3H^2$), which is also responsible for the exponentially vanishing in time of the phantom-like bulk scalar field.

Thus, we explicitly quote the non-trivial components of the projected to the brane non-local Weyl tensor 
\cite{maartenskoyama,SMS}:
\begin{eqnarray}
{\cal E}_{00} &=& \frac{3}{8}\left(\ddot u^2 - u'^2\right)~, \\
{\cal E}_{11} &=& {\cal E}_{22} = -\frac{e^{2Ht}e^{u}}{24}\left( 8u''- 3u'^2 + \right. \nonumber \\
&+& \left. 4\ddot u-3\dot u^2+12H\dot u \right)~, \label{E11}\\
{\cal E}_{33} &=& \frac{e^{2Ht}e^{-2u}}{24} \left( 16u''-15u'^2+8\ddot u + \right. \nonumber \\
&+& \left. 3\dot u^2+24H\dot u\right)~,
\label{E22}
\end{eqnarray}
where we have taken into account the relation (\ref{a}). In order to analyze the large time evolution of these 
quantities we shall focus on the exponential time dependence of (\ref{E11}) and (\ref{E22}). By considering the 
equations (\ref{separation})-(\ref{e}) in the expressions (\ref{E11}) and (\ref{E22}) we have
\begin{eqnarray} \label{E11t}
{\cal E}_{11} &=& -\frac{e^{2Ht}e^u}{8} \left[4e^{-3A/2}\left(3A'\chi - 2A'\chi' -
\Omega^2\chi\right)\epsilon\right.-\nonumber \\
&-& \left.e^{-3A}\left(\chi'-\frac 32 A'\chi\right)^2 \epsilon^2 - e^{-3A}\chi^2\ \dot\epsilon^2\right]~, \\
{\cal E}_{33} &=& \frac{e^{2Ht}e^{-2u}}{8}\left[8e^{-3A/2}\left(3A'\chi-2A'\chi'-\Omega^2\chi\right)\epsilon
-\right.\nonumber \\
&-& \left. 5e^{-3A}\left(\chi'-\frac 32 A'\chi\right)^2 \epsilon^2 + e^{-3A}\chi^2\dot\epsilon^2\right]~.
\label{E22t}
\end{eqnarray}
The common factors $e^u$ and $e^{-2u}$ super-exponentially become a unity with time in comparison with $e^{2Ht}$ and 
the expression in brackets and, hence, are not relevant for our analysis. We further note that for the under-damped 
and critical dumping cases, the qualitative behaviour of $\epsilon$ and its time derivatives for large times (up to 
factors of $\sin(\omega t+\delta)$, $\alpha t+\beta$ and their time derivatives) is governed by a decaying in time 
exponential factor that goes like
\begin{equation}
\epsilon \sim \dot\epsilon \sim \ddot\epsilon \sim e^{-3Ht/2}~,
\end{equation}
whereas the for the over-damped case their behaviour is dominated by a different decaying in time exponential factor 
of the form:
\begin{equation}
\epsilon\sim\dot\epsilon\sim\ddot\epsilon\sim e^{ \left(\tilde\omega - 3H/2\right)t}~,
\end{equation}
where we have taken into account that $3H/2 > \tilde\omega$ in accordance with (\ref{overe}) and (\ref{omegat}), 
hence, $e^{-3Ht/2}$ decays to zero faster than $e^{\tilde\omega t}$ grows.

Thus, for the under-damped and critical dumping cases, the second and third terms in the expression within the 
brackets of (\ref{E11t}) and (\ref{E22t}) always decay to zero faster than the overall factor $e^{2Ht}$ increases 
for large times. However, the first term within the brackets of these expressions is proportional just to 
$e^{-3Ht/2}$, and when multiplied by $e^{2Ht}$ leads to a total factor $e^{Ht/2}$ that eternally increases with time. 
Since the coefficients of the first terms in (\ref{E11t}) and (\ref{E22t}) are different, as time goes on, the 
components of the projected to the brane Weyl tensor (\ref{E11}) and (\ref{E22}) become more different, yielding a 
more and more anisotropic bulk as stated above.

The situation becomes even clearer in the over-damped case since the total factor that governs the large time 
behavior of (\ref{E11}) and (\ref{E22}) goes like $e^{\left(\frac{H}{2}+\tilde\omega\right)t}$ and yields an even 
more anisotropic bulk in comparison to the previous cases.


\section{Conclusions and discussion}

In this paper we have presented a novel mechanism of isotropization of an initially anisotropic 5D thick braneworld 
generated by a phantom-like scalar field minimally coupled to gravity. Under a wide class of suitable initial 
conditions (for any $t_i>0$ and arbitrary constants of the explicit scalar-tensor braneworld solution) the 
anisotropic braneworld evolves and super-exponentially isotropizes by itself, under the action of the 4D cosmological 
constant, while the scalar field exponentially disappears. Thus, the anisotropic energy of the 3-brane rapidly 
leaks into the bulk through the nontrivial components of the projected to the brane non-local Weyl tensor, leading 
to a less isotropic bulk. As a result we have an isotropic thick braneworld supported by pure curvature, where 4D 
gravity as well as other matter fields have been shown to be localized, the effective 4D Planck mass is finite and 
depending on the Hubble parameter, the 4D cosmological constant can adopt an infinitely small value without any 
fine-tuning since it is independent of the 5D one, and the mass spectrum of KK gravitons displays a mass gap, an 
important feature from the phenomenological point of view; moreover, the corresponding corrections to Newton's law 
have also been computed \cite{cuco,ghlmm}.

An interesting and puzzling point of this model concerns its stability. The usual unstable problems of phantom 
fields were tried to be avoided by leaving the scalar field to exist in the bulk but not in the branes, where it 
should couple to ordinary matter. However, both at the classical and quantum levels the scalar and metric 
fluctuations are strongly coupled and the stability issue of the brane system will remain still open. Notwithstanding, 
the scalar and its couplings have a very short life since they exponentially decay. In principle this is in agreement 
with the fact that the evolution of the universe experienced a phase transition from a highly anisotropic state to an 
isotropic one and, in this sense, it could have an initially unstable phase rapidly (super-exponentially) isotropizing 
to a stable de Sitter universe. This argument is reinforced by the fact that the anisotropic initial state of the 
universe was characterized by a Hubble parameter which is very large in comparison to its actual value, assisting a 
super-exponential isotropization which leads to an expanding universe of de Sitter type, i.e., in an accelerating way. 
The stability properties of the resulting isotropic braneworld are much simpler since there is no scalar field anymore 
and they were already studied in \cite{cuco}, where it was shown that the spectrum of tensorial metric fluctuations is 
free of tachyonic modes with $m^2<0$.

However, it should be pointed out as well that, as in other braneworld models, after the isotropization mechanism has 
occurred, the universe on the brane is still inflating under the action of the 4D cosmological constant, leading to a 
period of eternal inflation. This picture is different from the cosmological standard model where the isotropization 
of the 4D universe takes place thanks to the the fact that the inflaton field reaches its minimum after slow-rolling 
in a sufficiently flat self-interaction potential, giving place further to the phase of structure formation. It would 
be interesting to propose a richer model in the sense of considering standard matter on the brane that could lead to 
a more realistic model of the isotropization mechanism of the early universe.


\section*{Acknowledgements}
MG was supported by the grant of Rustaveli National Science Foundation $DI/8/6-100/12$ and by the research
project CONACYT $60060-J$. AHA is grateful to the staff of the ICF, UNAM for hospitality. AHA and UN thank SNI
for support. DMM acknowledges a postdoctoral fellowship from DGAPA-UNAM while RRML acknowledges a PhD grant from
UMSNH. This research was supported by grants CIC-UMSNH, CONACYT $60060-J$ and PAPIIT-UNAM, No. IN103413-3, 
{\it Teor\'ias de Kaluza-Klein, inflaci\'on y perturbaciones gravitacionales.}


\end{document}